\documentclass[12pt]{iopart}
\pdfoutput=1
\usepackage{color,graphicx,setspace,iopams,subfig,multirow}

\begin{document}

%opening
\title{Nanoscale broadband transmission lines for spin qubit control}
\author{J P Dehollain, J J Pla, E Siew, K Y Tan\footnote{Present address: Department of Applied Physics/COMP, POB 13500, 00076 AALTO, Finland}, A S Dzurak and A Morello}
\address{Centre for Quantum Computation and Communication Technology, School of Electrical Engineering and Telecommunications, University of New South Wales, Sydney NSW 2052, Australia}
\ead{jpd@unsw.edu.au}

\begin{abstract}
The intense interest in spin-based quantum information processing has caused an increasing overlap between the two traditionally distinct disciplines of magnetic resonance and nanotechnology. In this work we discuss rigourous design guidelines to integrate microwave circuits with charge-sensitive nanostructures, and describe how to simulate such structures accurately and efficiently. We present a new design for an on-chip, broadband, nanoscale microwave line that optimizes the magnetic field used to drive a spin based quantum bit (or qubit), while minimizing the disturbance to a nearby charge sensor. This new structure was successfully employed in a single-spin qubit experiment, and shows that the simulations accurately predict the magnetic field values even at frequencies as high as 30 GHz.
\end{abstract}

% Spintronics; Striplines; Semiconductor-device characterization, design, and modelling

\submitto{\NT}

\section{Introduction}\label{sec:intro} %%%%%%%%%%%%%%%%%%%%%%%%%%%%%%%%%%%%%%%%%%%%%

Magnetic resonance has been an essential tool to
study the physical and chemical properties of matter for over 60 years~\cite{slichter90}. For most of its history, it has been a ``bulk''
technique~\cite{fukushima81}, where a large ensemble of spins is
excited by means of radiofrequency and microwave excitations
delivered by a centimetre-size resonant structure, designed
according to microwave engineering guidelines~\cite{pozar97}. The scope of applications of magnetic resonance has evolved radically
since the realization that both electron~\cite{loss98PRA} and
nuclear~\cite{kane98nat} spins in the solid state can be used as qubits
for quantum information processing~\cite{ladd10N}. For this purpose,
the focus must shift towards the control and detection of individual
spins at the nanometre scale. This new and exciting line of research
has driven the fusion of two traditionally separate disciplines: magnetic resonance and nanotechnology.

Applying high-frequency oscillating magnetic fields to a single spin
is not any harder than to a large ensemble. The difficulty arises in
integrating the microwave excitation with the ultra-sensitive
detection techniques that must be employed to observe the signal of
a single spin. Several examples of successful single-spin resonance
in nanostructures exist. These employ a wide variety of techniques
to deliver the excitation to the spin, ranging from small coils~\cite{rugar04N,jelezko04PRL} and waveguides~\cite{xiao04N} to
nanofabricated coplanar striplines (CPS)~\cite{koppens06nat} and coplanar
waveguides (CPW)~\cite{fuchs09S}. These techniques tend to work well at relatively low frequencies ($< 1$~GHz) or
when the spin under study and the method used to detect it are
reasonably robust against heating and stray electric fields. A few
groups~\cite{simovic06RSI,obata07RSI} have discussed the challenges
they faced in integrating high-frequency ($>10$~GHz) microwaves with
single spins confined in GaAs quantum dots.

\begin{figure}
\begin{center}
\includegraphics[width=0.5\textwidth]{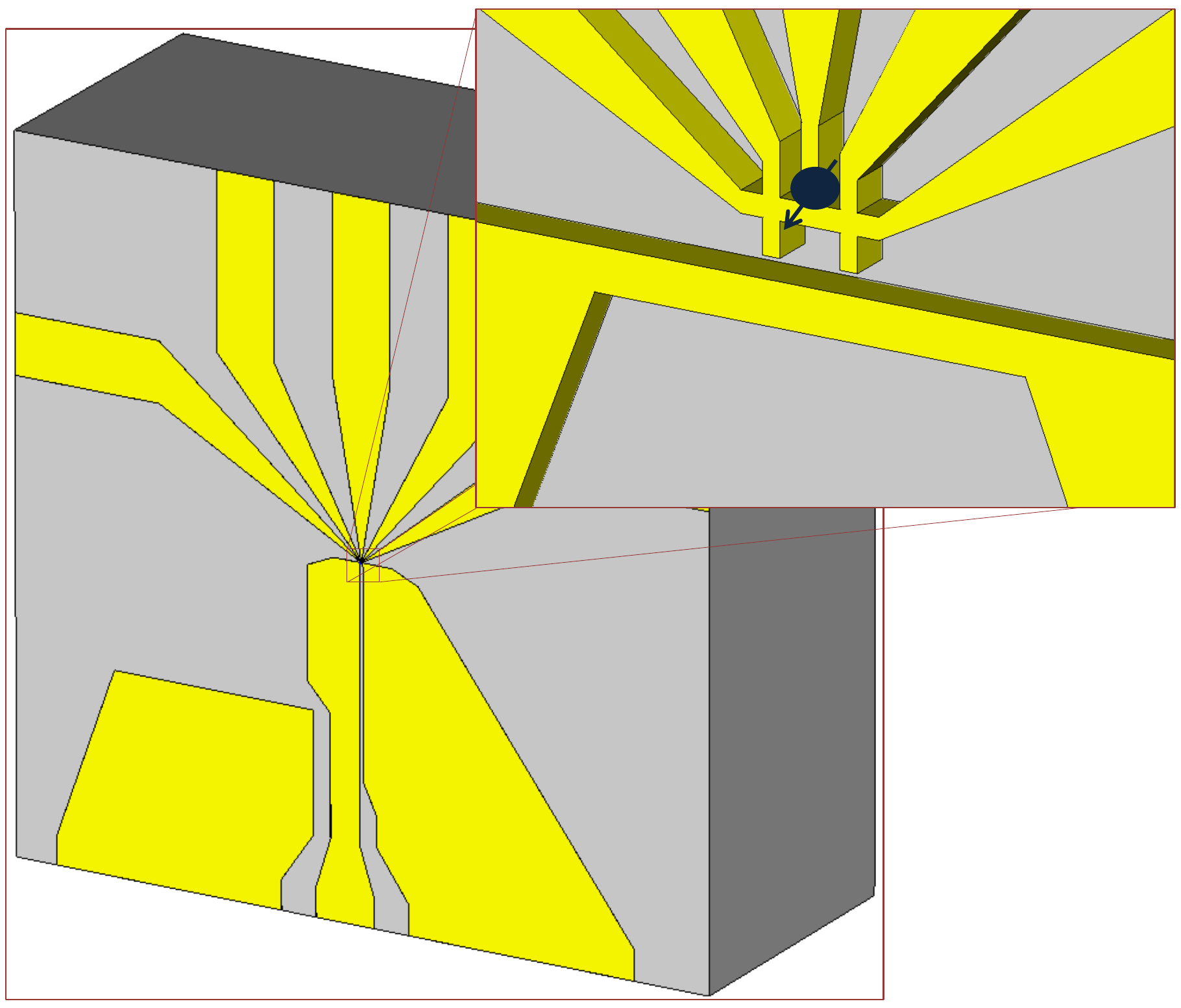}
\end{center} \caption{Optimized design for an on-chip broadband microwave transmission line for the control of a spin qubit. The inset shows a close-up of the nanoscale part of the structure, consisting of a short-circuit termination of the transmission line and a spin readout device. The dark sphere with arrow indicates the assumed location of the spin qubit. Alternative transmission line designs will be analyzed, while maintaining the same spin readout device structure.}\label{fig:Model}
\end{figure}

The aim of our work is to explain and assess a set of useful design
rules that can be applied to maximize the chance of success of
high-frequency spin resonance experiments in nanostructures. We illustrate the pros and cons of different designs by taking as an
example the architecture that was proposed for the control and
readout of single donor spins~\cite{morello09prb}, coupled to a
silicon Single-Electron Transistor (SET) for spin readout~\cite{morello10nat}. We
start with the broad microwave engineering framework to design a
planar transmission line terminating in a loop close to the spin qubit. Then we present a novel on-chip
transmission line for broadband localized $B_1$ field generation
(\fref{fig:Model}). We describe a methodology for carrying out
electro-magnetic field simulations on structures with large
dimensional range, from millimetre down to nanometre size. We then
use this tool to test the proposed transmission line and compare its
performance with other designs. Finally, we present experimental
measurements of our transmission line design and setup, and compare
with simulations.

\section{Design guidelines for the optimal planar loop}\label{sec:Guidelines} %%%%%%%%%%%%%%%%%%%%%%%%%%%%%%%%%%%%%%%%%%%%%

The application we have in mind for the planar transmission line and loop is to drive the spin resonance of a single electron, while the spin state is read out through an energy--- and spin-dependent tunnelling mechanism~\cite{elzermanN04,morello10nat}. The constraints posed by this type of application can be summarized as follows:

1 - The electron spin Zeeman splitting $E_Z = \hbar \gamma B_0$ must be larger than the thermal broadening $\sim 5k_B T$ of an electron reservoir at temperature $T$.   Here $\gamma = g \mu_B/\hbar$ is the gyromagnetic ratio, $g$ is the g-factor ($g \approx 2$ in Si, or $g \approx -0.4$ in GaAs),
$\mu_B$ is the Bohr Magneton, and $B_0$ is an externally applied static magnetic field. The corresponding resonance frequency, i.e. the frequency that the transmission line must deliver to the nanostructure, becomes $\nu_0 = E_Z / h$. For example, an electron temperature $T \sim 200$~mK would require a minimum resonance frequency $\nu_0 \sim 20$~GHz.

2 - The amplitude of the oscillating magnetic field $B_1$ produced by the loop should be maximized, to
allow fast rotations of the spin. An electron spin will flip (i.e. undergo
a $\pi$-rotation) in a time given by $t_{\pi} = h/(2g\mu_BB_1)$. The spin flip time
$t_{\pi}$ should be much shorter than the dephasing time $T_2^\star$ to ensure high-fidelity spin rotations.  $T_2^\star$ is obtained from Ramsey or free induction decay experiments, yielding e.g. $T_2^\star \sim 50$~ns in natural Si~\cite{lu11prb,pla12nat}. Equivalently, $B_1$ should be larger than the inhomogeneous width $\sigma_{\rm ESR}$ of the electron spin resonance line, typically due to the coupling of the electron to the bath of surrounding nuclear spins. Typical values of $\sigma_{\rm ESR}$ are $\approx 0.12$~mT in natural Si~\cite{tyryshkinPRB03}, and $\approx 2.2$~mT in GaAs~\cite{koppens06nat}. We note that, for the purpose of calculating the $B_1$ value relevant to spin resonance, the rotating wave approximation usually holds. A linearly polarized oscillating field should be decomposed into two counter-rotating components, only one of which contributes to the spin rotation. Therefore, the $B_1$ values obtained from a microwave simulator must be halved for the purpose of calculating e.g. a Rabi frequency.

3 - Electric fields radiated from the transmission line should be
minimized at the location of the spin qubit and readout device, since they can lead to unwanted effects
such as photon-assisted tunnelling~\cite{koppens06nat}, disrupt the operation of charge sensing devices (see section~\ref{sec:EField}), and contribute to the local heating of the nanostructure.

Given the constraints above, we will focus on simulations and design guidelines appropriate for spin resonance experiments in the 20--60~GHz range. However we note that a broadband planar loop that works well at $\nu > 20$~GHz will also exhibit good performance at MHz frequencies, since the losses and mode mismatches become less critical as the frequency is lowered. This is of relevance to the case where both electron spin resonance ($\nu > 20$~GHz) and nuclear magnetic resonance ($\nu \sim 100$~MHz) experiments can be performed on the same system, as in the case of dopant atoms~\cite{morton08nat}.

\subsection{Topologies of planar transmission lines}\label{sec:Designs} %%%%%%%%%%%%%%%%%%%%%%%%%%%%%%%%%%%%%%%%%%%

We consider here two main topologies of coplanar transmission lines: coplanar stripline and coplanar waveguide. The fundamental difference between the two is that the CPS is a \emph{balanced} line, meaning that it consists of two conductors, each having the same impedance to the surrounding ground planes. A CPS can carry microwaves in two modes: an \emph{odd} mode, where the potentials of the conductors oscillate in opposite phase, and an \emph{even} mode, where both conductors oscillate together with respect to the ground potential. The even mode is generally undesired, but can be excited in the presence of discontinuities and mismatches. The CPW instead is an \emph{unbalanced} line, consisting of a single conductor, while the ground planes act as return lines. It is also important to consider a planar transmission line topology known as microstrip, which consists of a single planar conductor, with a ground plane underneath the dielectric. Microstrip modes can be excited when coplanar lines are designed on a substrate that needs to sit on a conductive plate, and care needs to be taken to make sure the coplanar modes dominate the propagation of the signal.

Coaxial cables --- which we assume are going to be used to carry the microwave to the sample --- are unbalanced transmission lines. Therefore we analyze the most typical situation in which a coaxial cable is first coupled to a CPW fabricated on a printed circuit board (PCB). The CPW will have to be impedance-matched to the coaxial cable that delivers the microwave. The characteristic impedance of
planar transmission lines is mainly a function of the width and
thickness of the metal strips, the gap between coplanar strips and
the thickness of the dielectric. Various transmission line analysis textbooks~\cite{simons01,wadell91} provide equations to calculate the
impedance of many different planar transmission line types and
configurations.

\begin{figure}
\subfloat{\includegraphics[width=0.25\textwidth]{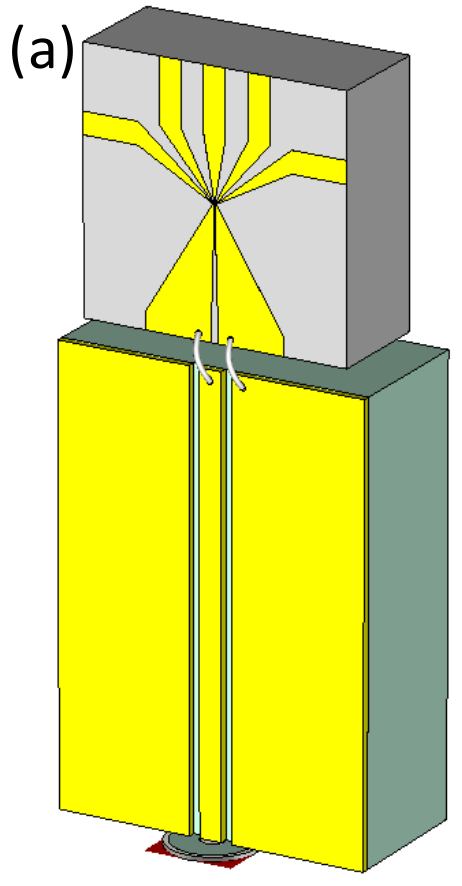}\label{fig:Simple}}
\subfloat{\includegraphics[width=0.25\textwidth]{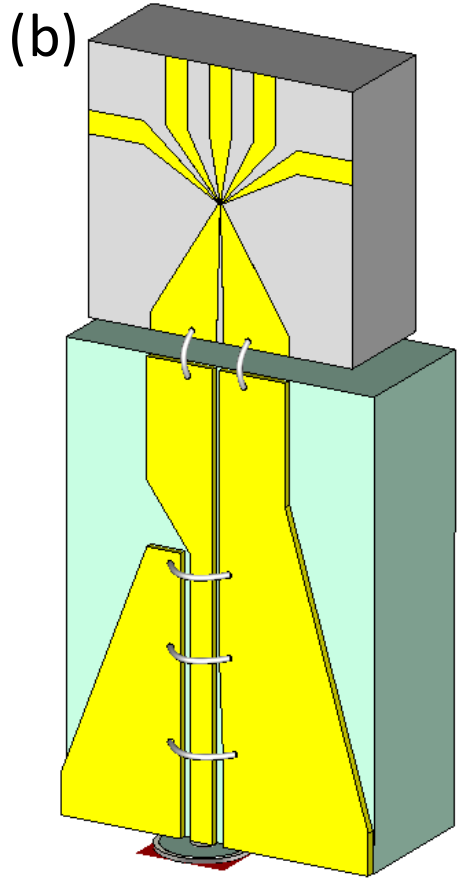}\label{fig:PCBalun}}
\subfloat{\includegraphics[width=0.25\textwidth]{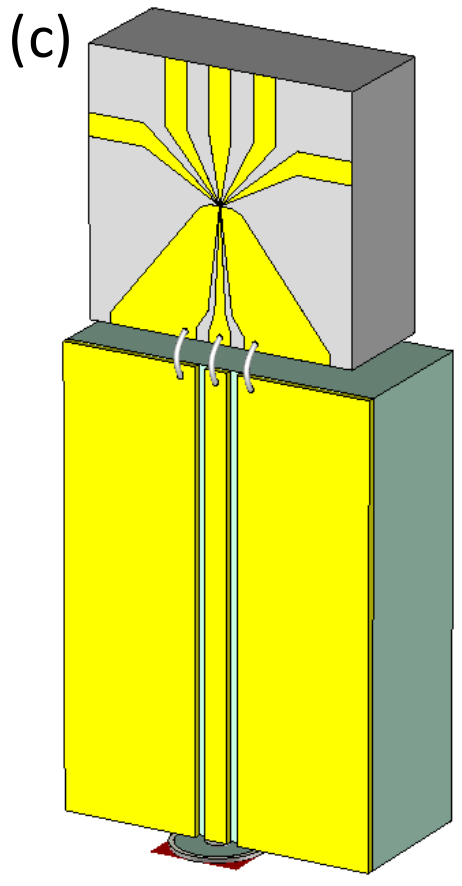}\label{fig:CPW}}
\subfloat{\includegraphics[width=0.25\textwidth]{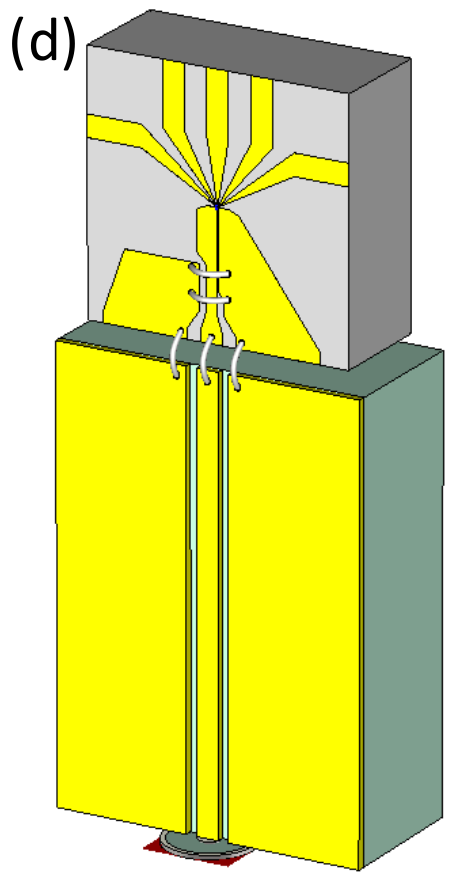}\label{fig:ChipBalun}}
\caption{Four planar loop designs. (a)~A simple CPW on a PCB directly bonded to an on-chip CPS. (b)~On-chip CPS with a CPW to CPS transition on PCB. (c)~CPW matched to the termination. (d)~Novel on-chip balun for maximized current at the loop.}\label{fig:Design}
\end{figure}

Referring to the drawings in \fref{fig:Design}, we analyze four possible configurations of lines and loops. Where a CPS is employed, we describe how to realize the conversion from unbalanced to balanced mode, known as a \emph{balun}.

\Fref{fig:Simple} is the simplest solution, consisting of a transition between the CPW on the board to a CPS on the chip, realized by bond wires. This design can match the line impedances, but it does not provide a well-controlled unbalanced to balanced line transition. The CPS is terminated by a short circuit next to the spin qubit, with the purpose of creating a node of the electric field --- to minimize the disturbance to the spin readout device (SRD) --- and an anti-node of the current --- to maximize the magnetic field that drives the spin resonance. Here it's crucial to recognize that the desired node of the electric field only occurs for a perfect odd-mode transmission. Accidentally exciting an even mode will cause an \emph{anti-node} of the electric field at the end of the line. This is a realistic danger for a CPW-CPS transition as crude as that shown in \fref{fig:Simple}. An on-chip CPS was used successfully by Koppens \emph{et al.}~\cite{koppens06nat} to drive the magnetic resonance of a single spin in a GaAs quantum dot at $\nu<1$~GHz, but contacting the CPS was done directly with a microwave probe instead of using bond wires.

\Fref{fig:PCBalun} shows an improved design, where a CPW-CPS balun is first fabricated on the PCB, then the CPS on the board is bonded to the CPS on the chip. Here the conversion to a balanced line occurs in a more controlled way as compared to Fig. \ref{fig:Simple}, and is less likely to excite an even mode of transmission. However, some risk still exists due to the use of bond wires to connect to the CPS on the chip. Planar baluns come in many forms and complexities. We chose the CPW to CPS transition presented by Chiou \emph{et al.}~\cite{chiou95el} and analyzed by Mao \emph{et al.}~\cite{mao00mtt}, which combines simple design, broadband operation and low insertion loss.

\Fref{fig:CPW} shows a design based solely on CPWs, both on the PCB and on the chip. The CPW on chip terminates with short circuits on both sides. This type of design was used for instance by Fuchs \emph{et al.}~\cite{fuchs09S} to drive the ESR of a nitrogen-vacancy centre in diamond at $\nu = 0.49$~GHz. Due to the absence of mode conversions, this design will have the widest range of operating frequencies and the lowest insertion loss. However, the nature of
the CPW short implies that the current is divided equally amongst
two loops, therefore only half of the signal received at the short
can be exploited to generate a magnetic field at the qubit location.

Finally, in \fref{fig:ChipBalun} we propose a novel planar loop design, which consists of a short-circuited, on-chip
CPW to CPS transition. In this design we keep
the matched CPW structure at the interface between PCB and chip, and make the
transition to CPS on the silicon chip. To achieve this, the
transition needs to be scaled to accommodate the limited dimensions.

\begin{table} \caption{Dimensions used in all the planar
transmission lines presented in this article. See \fref{fig:TLDim} for dimension definitions.} \begin{indented}
\lineup \item[]\begin{tabular}{@{}*{5}{l}} \br
\multicolumn{1}{c}{\multirow{2}[4]{*}{\textbf{Type of TL}}} &
\multicolumn{2}{c}{\textbf{Input}} &
\multicolumn{2}{c}{\textbf{Termination}} \cr \multicolumn{1}{c}{} &
\boldmath{}\textit{\textbf{$a$ ($\mu$m)}}\unboldmath{} &
\boldmath{}\textit{\textbf{$b$ ($\mu$m)}}\unboldmath{} &
\boldmath{}\textit{\textbf{$a$ ($\mu$m)}}\unboldmath{} &
\boldmath{}\textit{\textbf{$b$ ($\mu$m)}}\unboldmath{} \cr \mr CPW on
PCB&100&\060&100&\060\cr Balun on PCB&100&\060&\030&300\cr CPW on
chip&100&\060&\0\01.6&\0\01\cr CPS on chip&\030&300&\0\01&\010\cr
Balun on chip&100&\060&\0\07&\090\cr \br \end{tabular}%
\end{indented} \label{tab:TLDim}%
\end{table}%

\begin{figure}
\begin{center}
\includegraphics[width=0.5\textwidth]{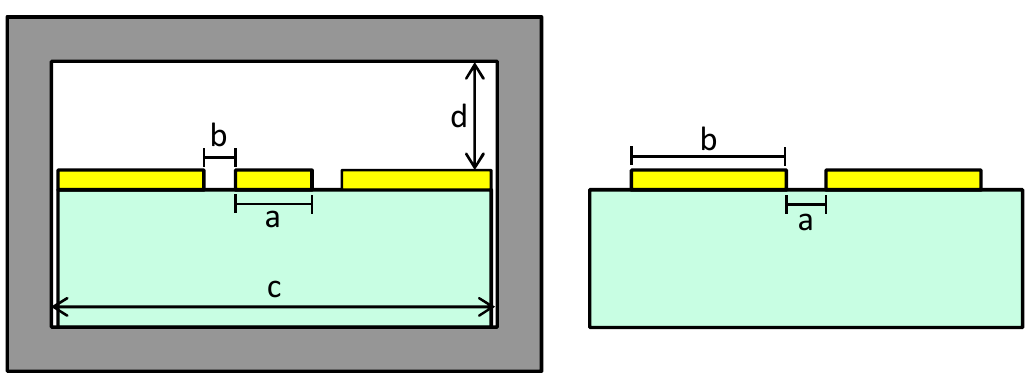}
\end{center}
\caption{Conventions used in this article for the dimensions of CPW (left), CPS (right) and enclosures (shown in CPW image). For CPW: $a$ is the width of the centre metallic strip, $b$ is the pitch between the signal strip and the ground plates. For CPS: $a$
is the pitch between the balanced strips, $b$
is the width of each metallic strip. For enclosures: $c$ is the width, $d$ is the height from the top the transmission line.}\label{fig:TLDim}
\end{figure}

In realistic experimental conditions, other parameters can influence the performance of the transmission lines. The chip will normally be mounted in a metallic enclosure for thermal and electromagnetic shielding. Delivering microwaves into the enclosure can excite cavity modes which can severely perturb the behaviour of the transmission lines. Therefore, the dimensions of the enclosure must be designed to minimize cavity modes. From microwave engineering textbooks~\cite{wadell91}, it is known that for CPW and CPS, keeping $c/(a+2b) > 1.75$ and $d/a > 2.5$ (see \fref{fig:TLDim}) ensures the enclosure will not affect the impedance of the transmission line by more than 1.5\%. Microstrip modes are minimized by making $a+2b$ shorter than the thickness of the dielectric.

The shape and length of the wire bonds connecting the PCB and chip
will also have an important effect on the signal losses at higher
frequencies. Bond wires have an effective inductance which increases
as they get longer and thinner. With the help of simulations, we
analyze this effect in more detail in \sref{sec:Optimization}.\\

With regards to the design of the balun included in the topologies in \fref{fig:PCBalun} and \fref{fig:ChipBalun}, we follow the method used by Mao \emph{et al.}~\cite{mao00mtt}, where the CPW to CPS transition is divided into five sections. Each section tapers, crops and/or expands
the transmission line to gradually make the transition keeping
impedance variations small and fluid. The length of each section and
the angle of the discontinuities needs to be optimized to minimize
losses. When scaling the transitions to fit our PCB and chip
dimensions, we ran a parametric sweep of simulations to optimize
each transition section. The discontinuities at each section of the
transition can excite the even mode of propagation, which can
increase losses in the transmission line. This effect can be
suppressed by connecting the ground planes with bond wires or air
bridges where the discontinuities occur.

\subsection{Choice of simulation methods for tapered structures} %%%%%%%%%%%%%%%%%%%%%%%%%%%%%%%%%%%%%%%%%%%%%

\begin{figure}
\includegraphics[width=\textwidth]{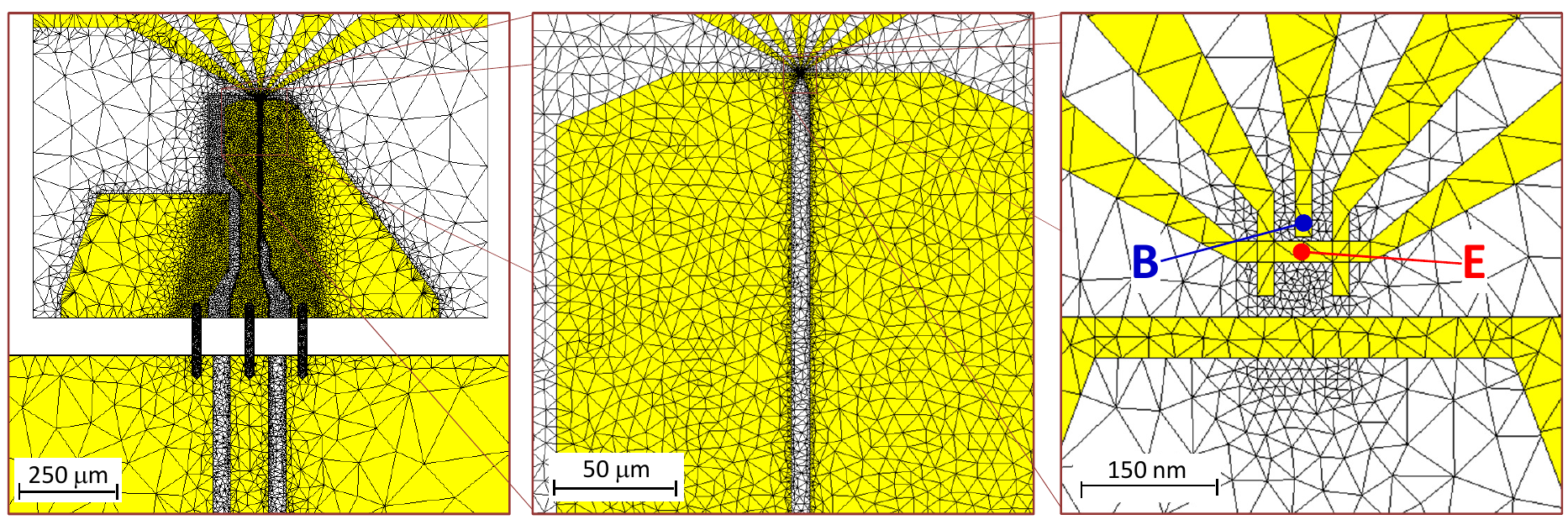} \caption{Tetrahedral mesh
used for simulations. Mesh density is increased sequentially,
adjusting to the sharp tapering of the structure. The rightmost inset shows the locations of the probes for all our simulation results: the magnetic field probe is at the donor site location (blue); the electric field probe is at the charge detector (red).}\label{fig:Mesh}
\end{figure}

Modelling and simulation of the structures described in this paper is carried out using the
Microwave Studio software package, from Computer Simulation Technology (CST-MWS)~\cite{cst}.
The choice of a suitable solver is based upon three requirements:

1 - In most spin qubit device structures, the spin is buried at some depth below the surface on which the planar loop is fabricated. Therefore the electromagnetic simulation --- the goal of which is to obtain the magnetic and electric field at the qubit location --- must be carried out with a volume-based method.

2 - We are interested in obtaining broadband results, since the operating frequency of the spin qubit might vary over a wide range, depending on the applied magnetic field.

3 - All the structures described in \fref{fig:Design} are sharply tapered, with edge lengths shrinking from
millimetres to nanometres. The electromagnetic solver must handle correctly such wide range of structures sizes.

The first requirement reduces our choice of available mesh types to
hexahedral and tetrahedral. CST-MWS uses the finite integration
method (FIM)~\cite{weiland77aeu} to support both these mesh types
with time-domain and frequency-domain solvers. The chosen method
should provide accurate results, in the least amount of simulation
time.

FIM in the time-domain is a leap frog algorithm similar to the
finite-difference time-domain method~\cite{yee66tap}. It is useful
for obtaining broadband results, as a single time domain simulation
can be transformed to obtain an arbitrarily wide frequency response.
However, this method requires a structured grid and only supports
hexahedral meshing in CST-MWS. This means that for an accurate
representation of our structure, the mesh around the small features
will generate a great amount of unneeded mesh points at the large
features. Simulation time for this scheme increases linearly on the
number of mesh points and the minimum distance between mesh points.

In the frequency-domain, FIM solves a set of Maxwell's equations for
the entire volume, in one simulation. This makes the algorithm
independent of the grid structure, which allows support for the more
flexible tetrahedral meshing. Simulation times only depend on the
total number of mesh points. The disadvantage of this scheme is that
each simulation solves for only one frequency point, therefore it is
necessary to run several simulations in order to obtain a broadband
result.

Attempting to accurately represent our tapered structure with
hexahedrons will result in an unmanageably large mesh. Therefore the
best option is to use the frequency domain solver with tetrahedral
meshing. The immense difference in simulation times makes it worth
running many simulations in the frequency domain, instead of one
simulation with the time domain solver using hexahedral meshing.

Once the solver and mesh type have been selected, the next key step
is to produce a mesh that adequately represents all the materials,
corners and edges of the structure, with the minimum number of mesh
points. CST-MWS can be configured to automatically create an initial
mesh and run a mesh adaptation algorithm. The software will
sequentially run a simulation and modify the mesh until a
convergence is reached on scattering parameters (S-parameters)
error. The algorithm adapts the mesh to increase its density at the
electrically relevant locations (i.e. corners and edges). However, the automated meshing scheme fails to accurately adapt the
mesh to accommodate the large feature size range in our sharply tapered structure. It is therefore necessary to manually
divide the model into different sections and assign different
meshing densities to each section (see \fref{fig:Mesh}). With the
mesh adjusted to both feature size and electrical relevance, we can
obtain accurate results with minimal mesh points.

Boundary conditions must be set for each of the planes that define
the limits of the modelling space. Boundary conditions can be
defined to either minimize reflections (\emph{open space}), or behave like
a perfect metallic object (\emph{electric}). To simplify simulations, we
model a smaller PCB than what would typically be used in an
experimental setup, and we set the boundary conditions to open space
on the front and lateral sides of the structure. On the back side the boundary condition is set to electric, to observe the effects of microstrip modes in our transmission lines. These boundary conditions give accurate results as long
the guidelines for enclosure size discussed in \sref{sec:Designs} are followed.

Electric and magnetic field probes can be placed anywhere in the three dimensional space of the structure. As shown in \fref{fig:Mesh}, the electric field probe is placed directly underneath the gate of the charge detector; while the magnetic field probe is placed at the donor site (i.e. where the spin to be controlled is located). We extract only the component of oscillating magnetic field that is perpendicular to the static magnetic field.

Weiland \emph{et al.}~\cite{weiland08mm} provides a more detailed explanation of the simulation parameters described in this section.\\

All the models presented in this article comprise a modelled coaxial
input port, a PCB with a planar transmission line, and a Si chip
with a planar loop next to a spin readout device similar to
the one used by Morello \emph{et al.}~\cite{morello10nat}. We always assume 1~mW (0 dBm) of
power at the coaxial input port of the simulated structure. We modelled the PCB using the
characteristics of the Rogers RO3010 laminate with a thickness of
640~$\mu$m and 35~$\mu$m copper (Cu) cladding. The planar
transmission line on the PCB bridges the coaxial port and the chip.
It is connected via bond wires to the planar loop, which is assumed to be fabricated
in 100~nm thick aluminium (Al), on top of a silicon (Si) chip with
a surface of 1.2$\times$1.2~mm$^2$ and a thickness of 500~$\mu$m. As shown in \Fref{fig:Model}, the loop is located at the centre of the chip surface, with a minimum width of 100~nm at the short circuit. The SRD is placed approximately 130~nm away from the loop. The conductivity of the Cu lines is increased by two orders of magnitude as compared to the room-temperature (RT) textbook values, to account for the use of the device at cryogenic temperatures~\cite{berman52prs,kumar93jms}. The conductivity of the Al lines is increased by a factor 3.85 as compared to its RT value, based on independent resistance measurements of a coplanar loop similar to the one described here, at RT and 4.2 K. We assume the structure is always placed in a magnetic field large enough to suppress the superconductivity in Al. The bonding pads of all Al strips have a minimum area of 100$\times$100~$\mu$m$^2$. Unless otherwise stated, the bond wires have 25~$\mu$m diameter and a length of approximately 200~$\mu$m. \Tref{tab:TLDim} shows the relevant dimensions of all the planar transmission line types we present.

\section{Simulation Results} %%%%%%%%%%%%%%%%%%%%%%%%%%%%%%%%%%%%%%%%%%%%%%%%%%%%%%

\subsection{Electric and magnetic fields along the transmission line} %%%%%%%%%%%%%%%%%%%%%%%%%%%%%%%%%%%%%%%%%%%%%

\begin{figure}
\subfloat{ \includegraphics[width=0.36\textwidth]{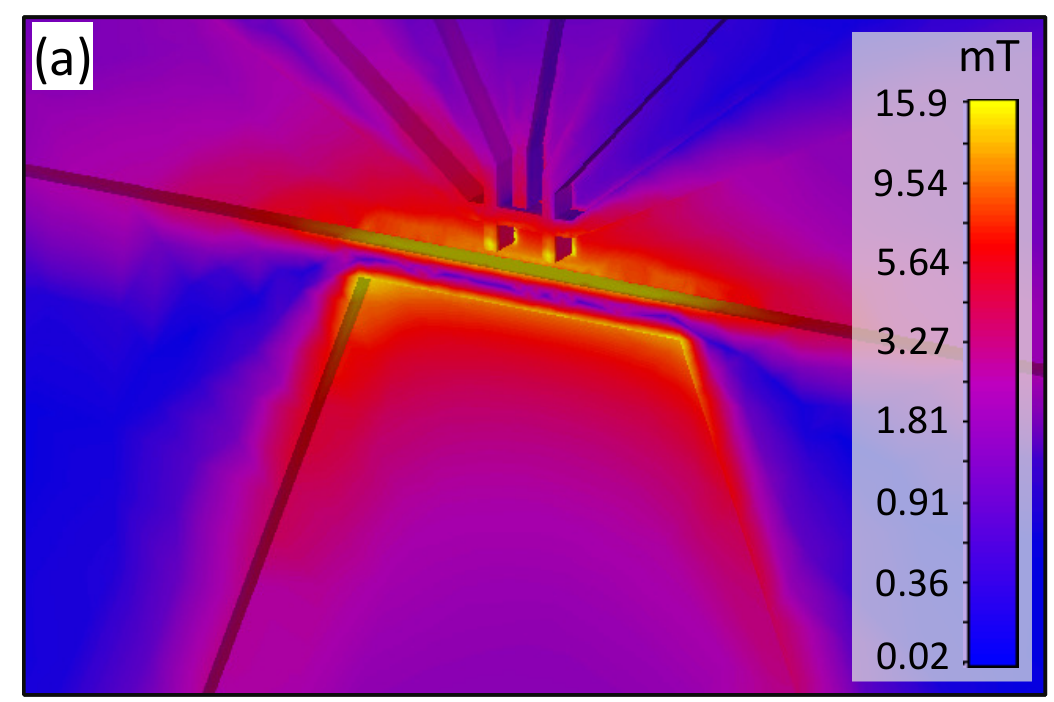} \label{fig:2DBfield}}
\subfloat{ \includegraphics[width=0.2\textwidth]{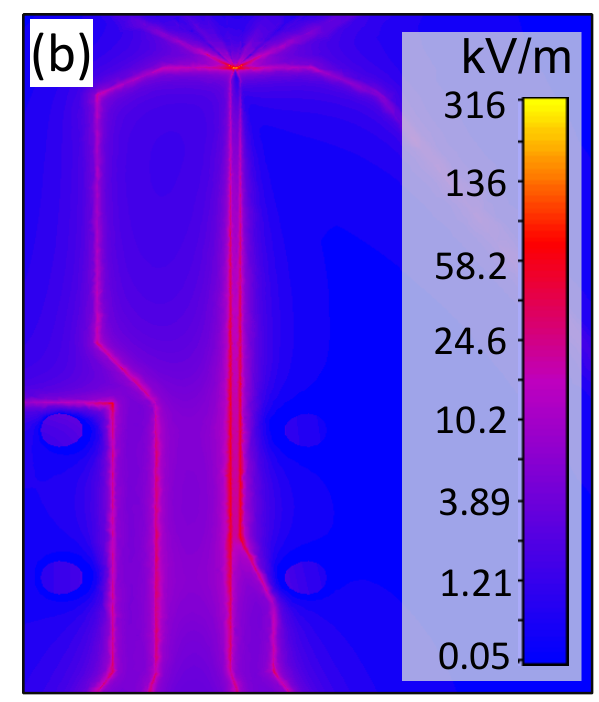} \label{fig:2DEfield}}
\subfloat{ \includegraphics[width=0.44\textwidth]{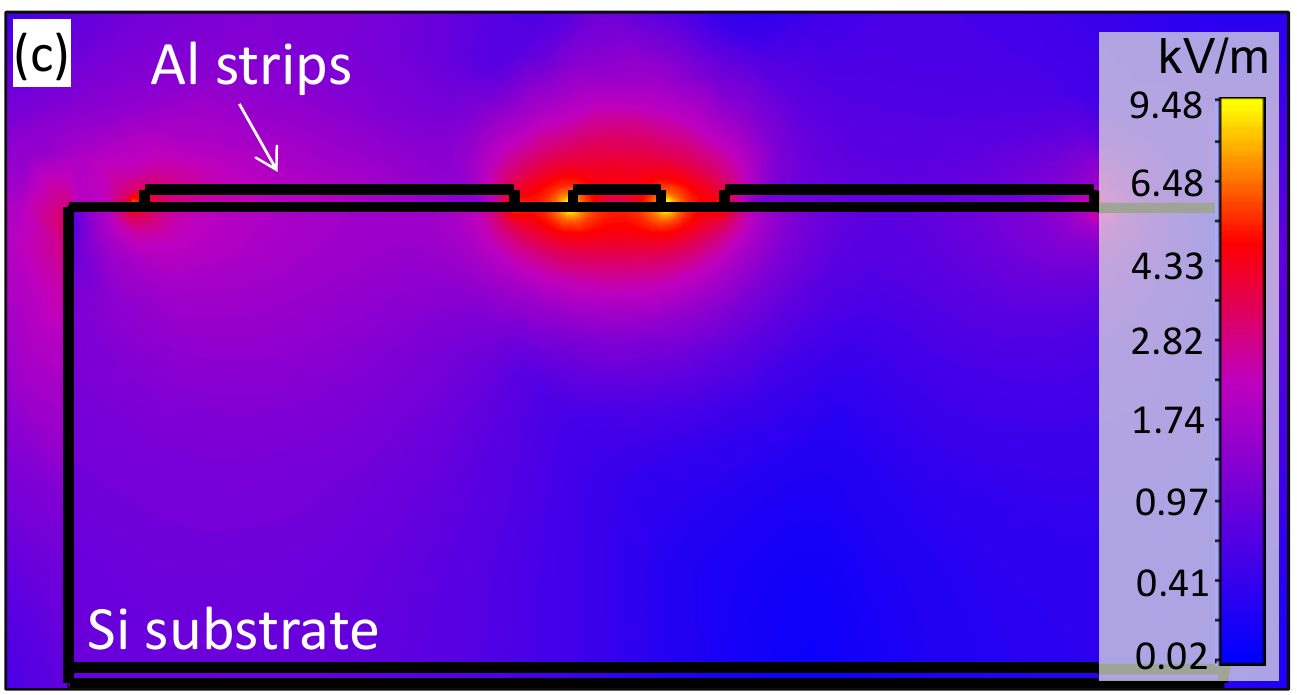} \label{fig:CPWmode}}
\caption{Characterization of the planar transmission lines. (a)~2D
plot of the component of the magnetic field amplitude perpendicular to the surface of the chip.
(b)~2D plot of the absolute value of the electric field. (c)~Electric field at the input of the on-chip CPW. The field is mainly radiated from the centre Al conductor to the coplanar ground planes. The fields in all figures are calculated at 50 GHz.}
\end{figure}

Using the simulation setup described in the previous section, we started by carrying out a characterization of the electric and magnetic field profiles along the short
circuited coplanar transmission line shown in \Fref{fig:Model}. \Fref{fig:2DBfield} shows the amplitude of the perpendicular component of the magnetic field $B_1$ generated around the loop. $B_1$ clearly decays
with distance from the short, which shows the importance of placing
the loop as close as possible to the spin qubit.

The voltage difference between signal and ground lines generates an
electric field, which decreases to zero (\fref{fig:2DEfield}) as it
approaches the voltage node at the short-circuit. However, the resonant stub behaviour of the on-chip line (see \sref{sec:Results}) causes an additional electric field to be emitted at the end of the line. This electric field can be responsible for a degradation of the performance of the SRD, and heating of the electron layer nearby.

\Fref{fig:CPWmode} shows that the transmission through the on-chip line is dominated by the CPW mode, as desired. The field being radiated from the centre conductor to the back plate is negligible compared to that radiated to the coplanar planes.

The metal gates of the spin readout device couple capacitively to the loop, causing an increase in the electric field radiated by the loop. Simulating our structure without and with a SRD shows an electric field increase of $\approx 1.5$ orders of magnitude over the whole frequency range. In contrast, the magnetic field only increases by $\approx 30 \%$ in the presence of the SRD.

The surface current monitor allows us to estimate the current flowing through the loop. We find that 80~$\mu$A of current flowing through the planar loop will generate 0.1~mT of magnetic field at the spin location.

The skin depth effect can reduce the effective conductivity of
transmission lines, therefore increasing signal loss in transmission. The skin depth $\delta$ can be calculated as
$\delta = \sqrt{2\rho/\omega\mu}$ where $\rho$ is the resistivity of
the conductor, $\omega$ is the angular frequency of the current and
$\mu$ is the absolute permeability of the conductor. The fields in the conductor decay by an amount proportional to $1/e^\delta$ \cite{pozar97}. Therefore having a conductor of thickness $> 4\delta$ ensures more than 98 \% of the field has decayed, with no significant reduction in the effective conductivity of the line. Between 20 and 60~GHz, the skin depth of an aluminium line at cryogenic
temperatures ($\rho = 7$~n$\Omega$/m) is 172~nm to 300~nm. The on-chip transmission lines modelled have conductor widths that comfortably exceed the skin depth limits, however the 100 nm thickness of the Al clad could potentially affect the transmission. Results from simulations of a similar model with a 1 $\mu$m thick on-chip transmission line, show there is no appreciable difference in transmission parameters or field amplitudes.

\subsection{Performance comparison between CPS/CPW topologies}\label{sec:Results} %%%%%%%%%%%%%%%%%%%%%%%%%%%%%%%%%%

\begin{figure}
  \begin{center}
  \subfloat{\includegraphics[width=0.7\textwidth]{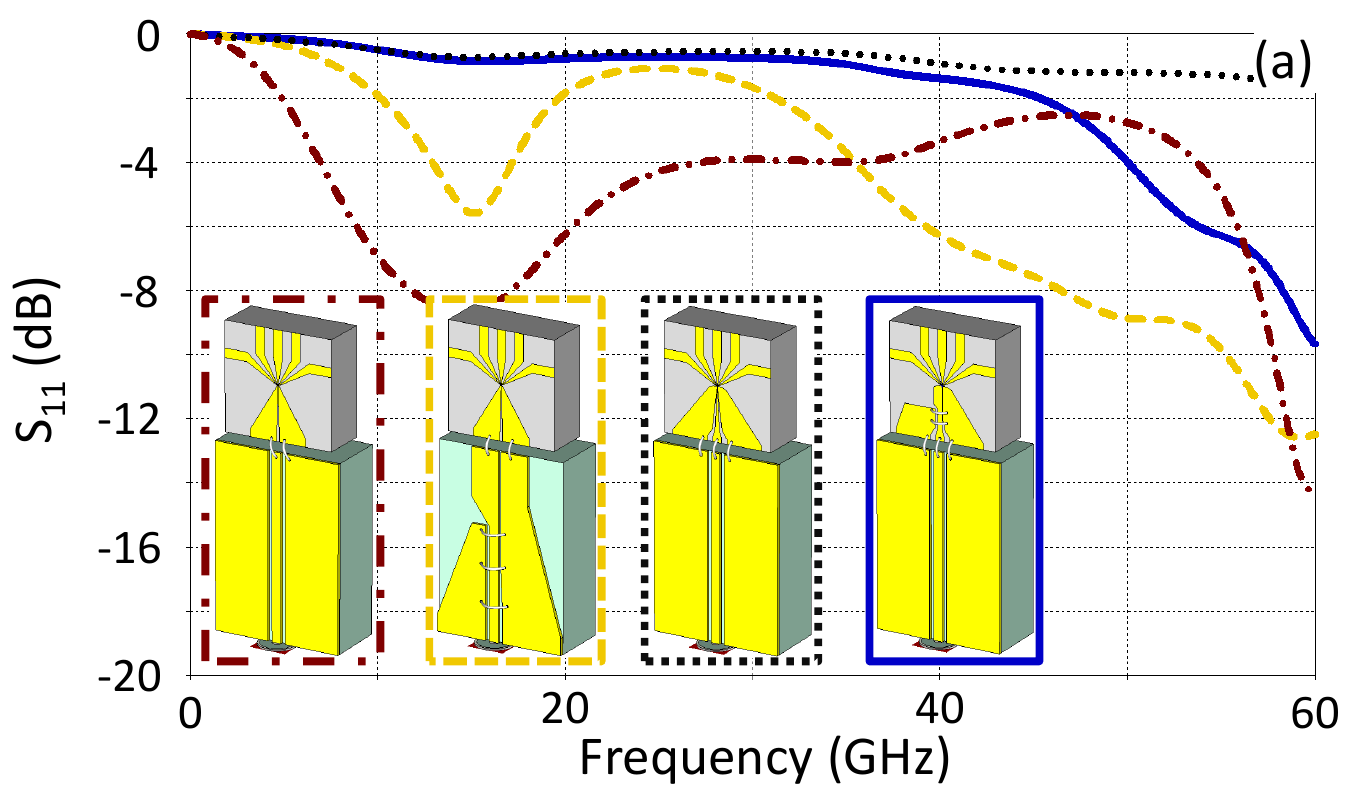}\label{fig:S11Sim}}\\
  \subfloat{\includegraphics[width=0.7\textwidth]{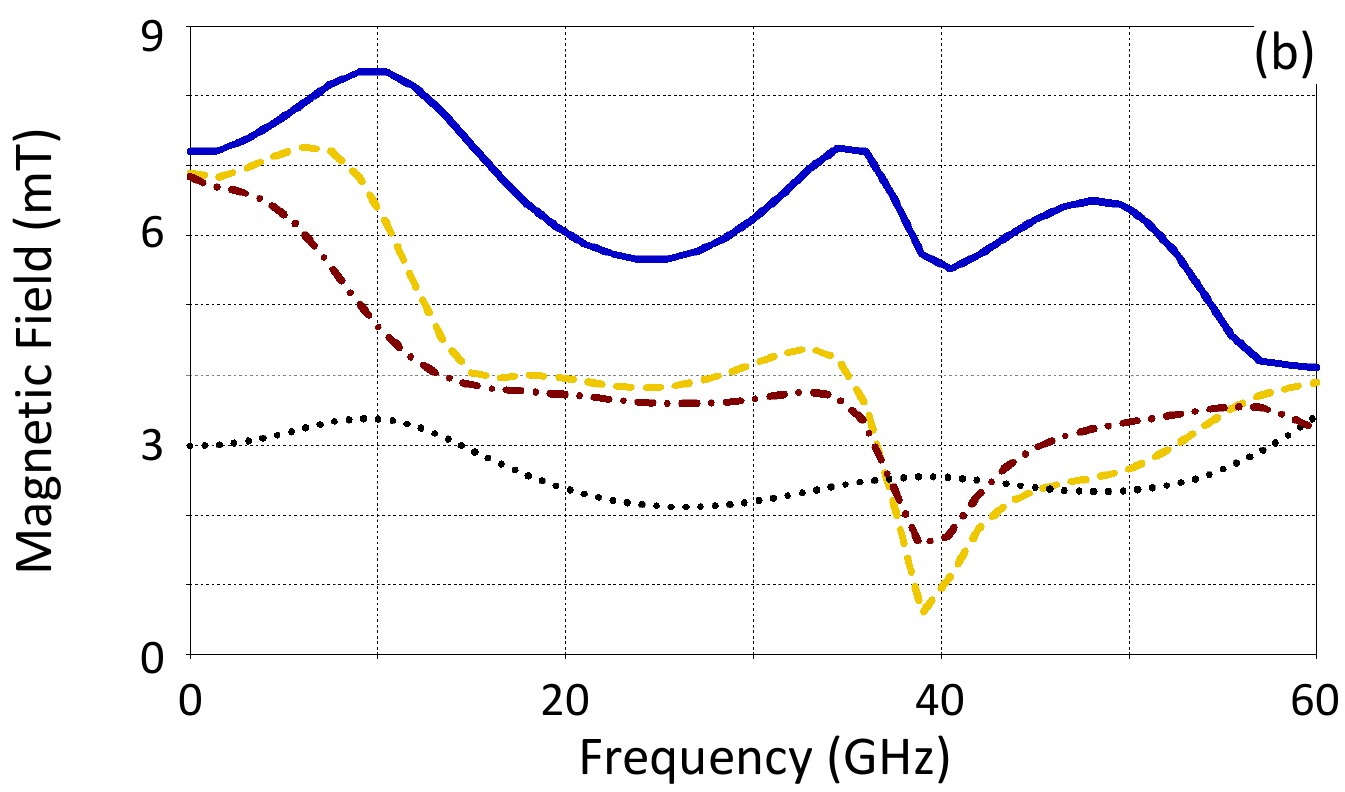}\label{fig:B1Sim}}\\
  \subfloat{\includegraphics[width=0.7\textwidth]{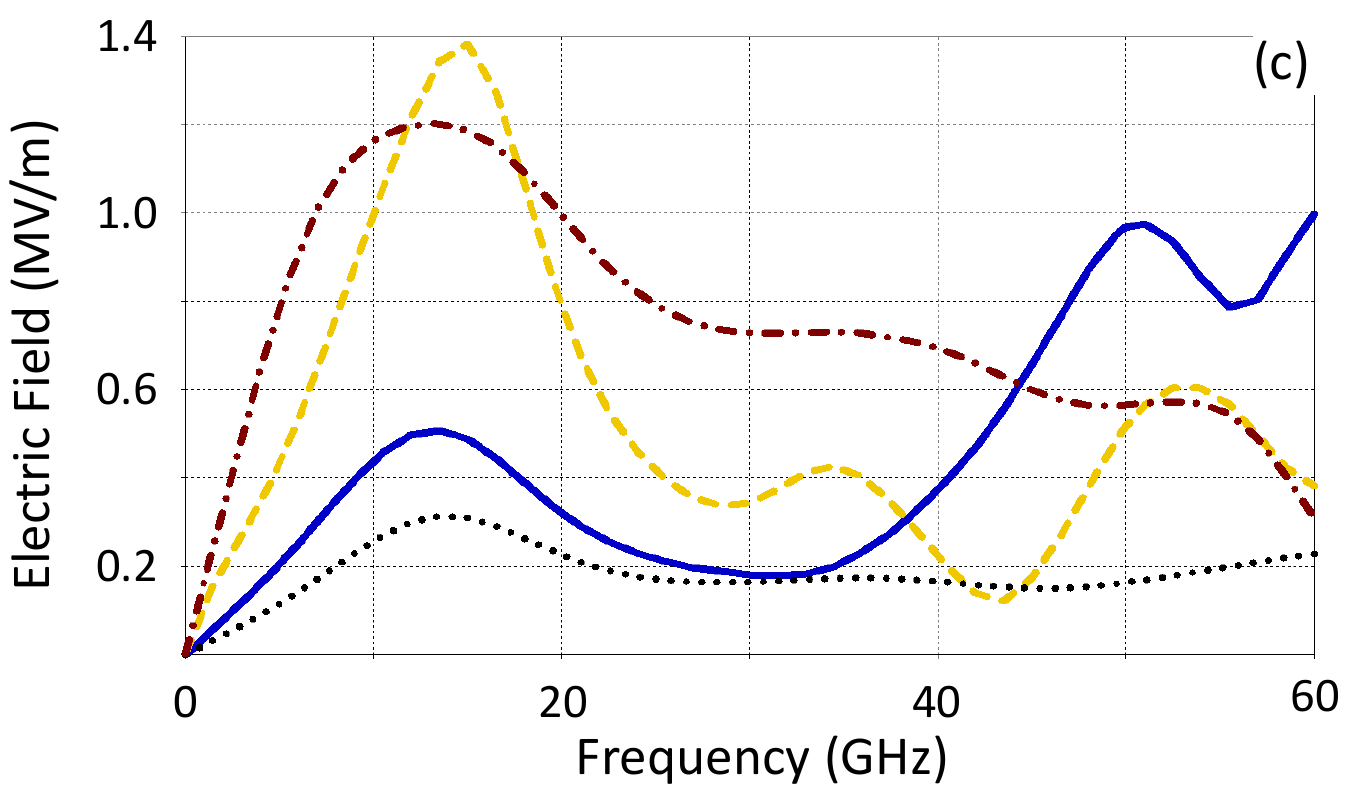}\label{fig:ESim}}
  \end{center} \caption{Simulation results comparing the frequency
  response of four planar loop designs: On-chip CPS with no
  transitions (\chain); On-chip CPS with PCB balun (\dashed);
  On-chip CPW (\dotted); On-chip balun (\full). (a)~S$_{11}$
  parameter. (b)~Magnitude of the perpendicular component of the
  magnetic field at the donor site. (c)~Absolute electric field at the
  charge detector, in this case the SET island.}\label{fig:SimResults}
\end{figure}

The four planar transmission line models presented in
section~\ref{sec:Designs} were simulated and the results are
compared in \fref{fig:SimResults}. The S$_{11}$ parameter, which represents the reflected power, is a useful measure of the broadband performance of the lines. An ideal
short-circuited transmission line should be fully reflective,
yielding S$_{11} = 0$~dB at all
frequencies. S$_{11} < 0$~dB indicates radiative losses, and we indeed observed that the behaviour of S$_{11}$ correlates with the electric field radiated by the line towards the SRD.

The simulations show that the fully-CPW structure shown in \fref{fig:CPW} is the closest to
having the ideal behaviour S$_{11} \approx 0$~dB from 0 to 60~GHz and, accordingly, the smallest electric field radiated at the SRD. This is not surprising, since this
design is matched in impedance and propagation throughout. However, as mentioned before, this comes at the price of roughly halving the strength of $B_1$ available at the spin qubit location.

The on-chip balun design of \fref{fig:ChipBalun} generates the highest $B_1$ fields
through most of the spectrum. It radiates adequately low electric
fields, comparable to the CPW at low frequencies, but shows a resonance
at around 60~GHz, where the electric field increases accordingly. This phenomenon, known as ``stub resonance'', is caused by the discontinuity between the PCB and the Si chip, which makes the on-chip section of the transmission line behave like a short-circuited resonator~\cite{pozar97}. A short-circuited
line acts as a parallel R-L-C circuit when its length equals a
quarter wavelength ($\lambda/4$). In this type of resonator, a $\lambda/4$ standing wave is formed by having a minimum of the current at the input and a maximum at the loop. The on-chip line modelled here has a length of 550~$\mu$m, thus the resonance should be observed at $\lambda =
$~2.2 mm. Frequency and wavelength in microwaves relate to each
other through the propagation velocity, given by $v =
c'/\sqrt{\epsilon_{eff}}$, where $\epsilon_{eff}$ is the effective
dielectric constant of the medium through which the wave is
travelling. Due to the fact that waves in planar lines travel at the
interface of two different media (in this case vacuum, $\epsilon_r = 1$, and silicon, $\epsilon_r = 11.9$), obtaining
$\epsilon_{eff}$ for our on-chip balun is non-trivial. Assuming an
ideal planar line, half of its field travels in the substrate
(filling factor of 50\%), and its effective dielectric constant is
the average between the two media. With this assumption we can estimate $\epsilon_{eff} \approx 6.45$, and a stub resonance at $f = v/\lambda = 53.7$~GHz. The stub resonance
observed in the simulations is shifted to a higher frequency due
to the low inductance of the bond wires interfacing the PCB and
chip. In \sref{sec:Optimization}, these effects are analyzed in
further detail.

The design with propagation and mode mismatch (\fref{fig:Simple}) shows the poorest
performance at high frequencies, radiating more than two times the amount of electric field compared to other designs.

\subsection{Further optimization}\label{sec:Optimization} %%%%%%%%%%%%%%%%%%%%%%%%%%%%%%%%%%%%%%%%%%%%%

\begin{figure}
\begin{center}
\subfloat{\includegraphics[width=0.5\textwidth]{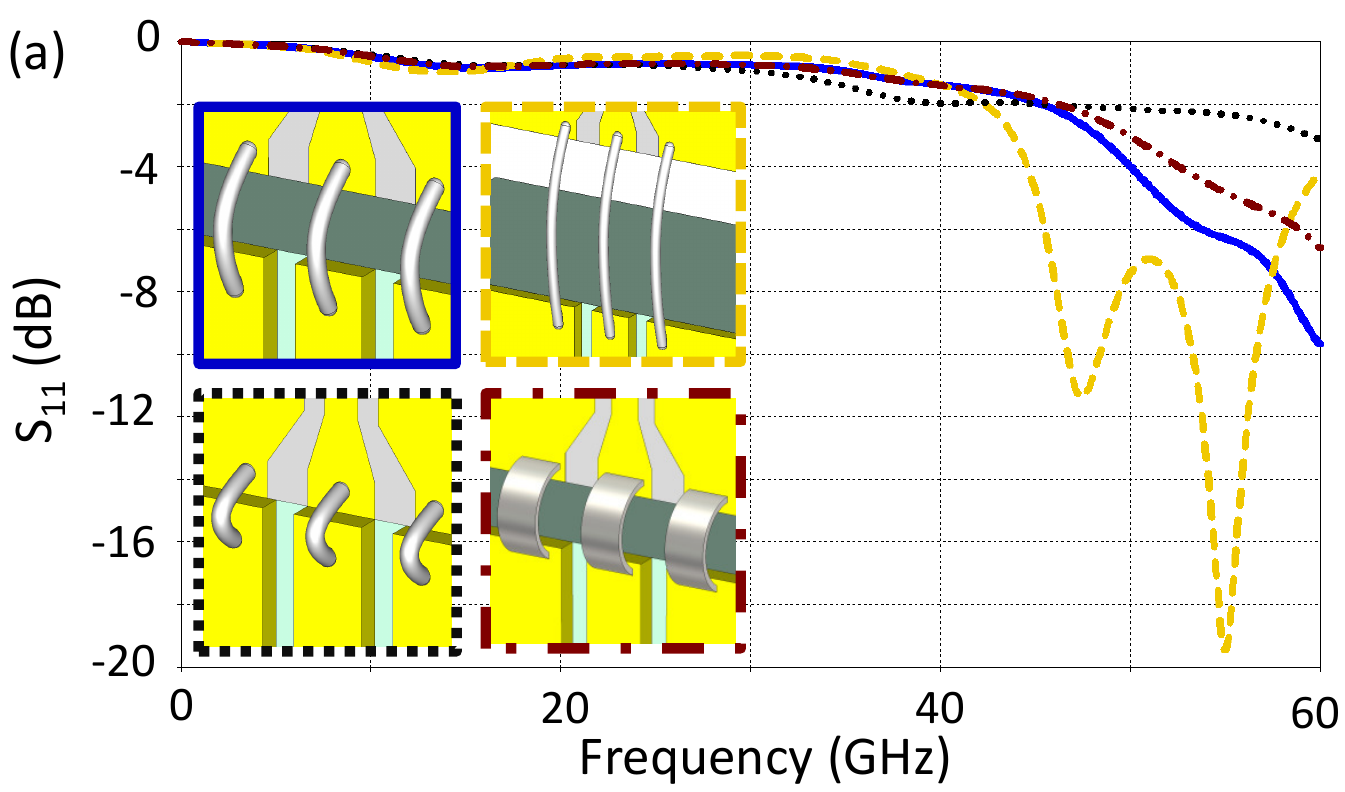}}\\
\subfloat{\includegraphics[width=0.5\textwidth]{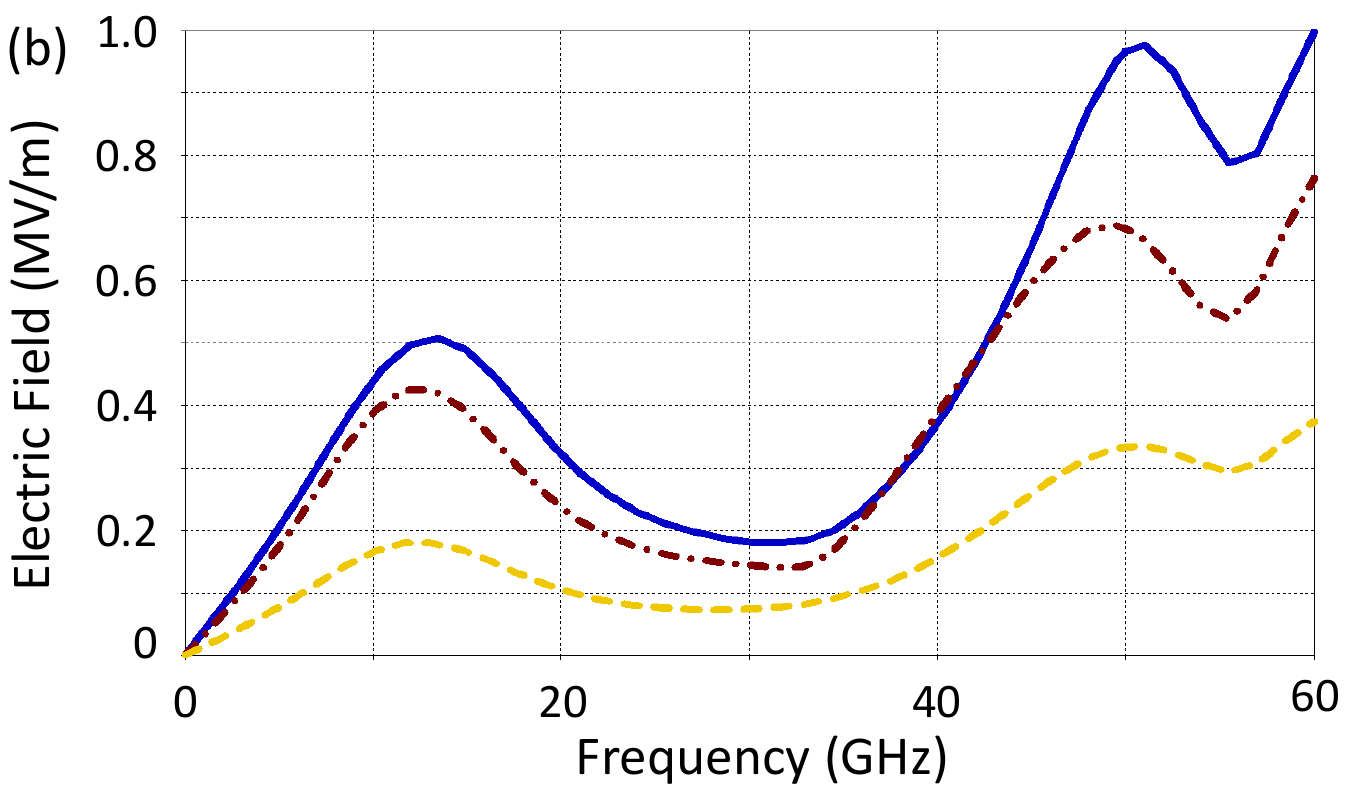}}
\subfloat{\includegraphics[width=0.5\textwidth]{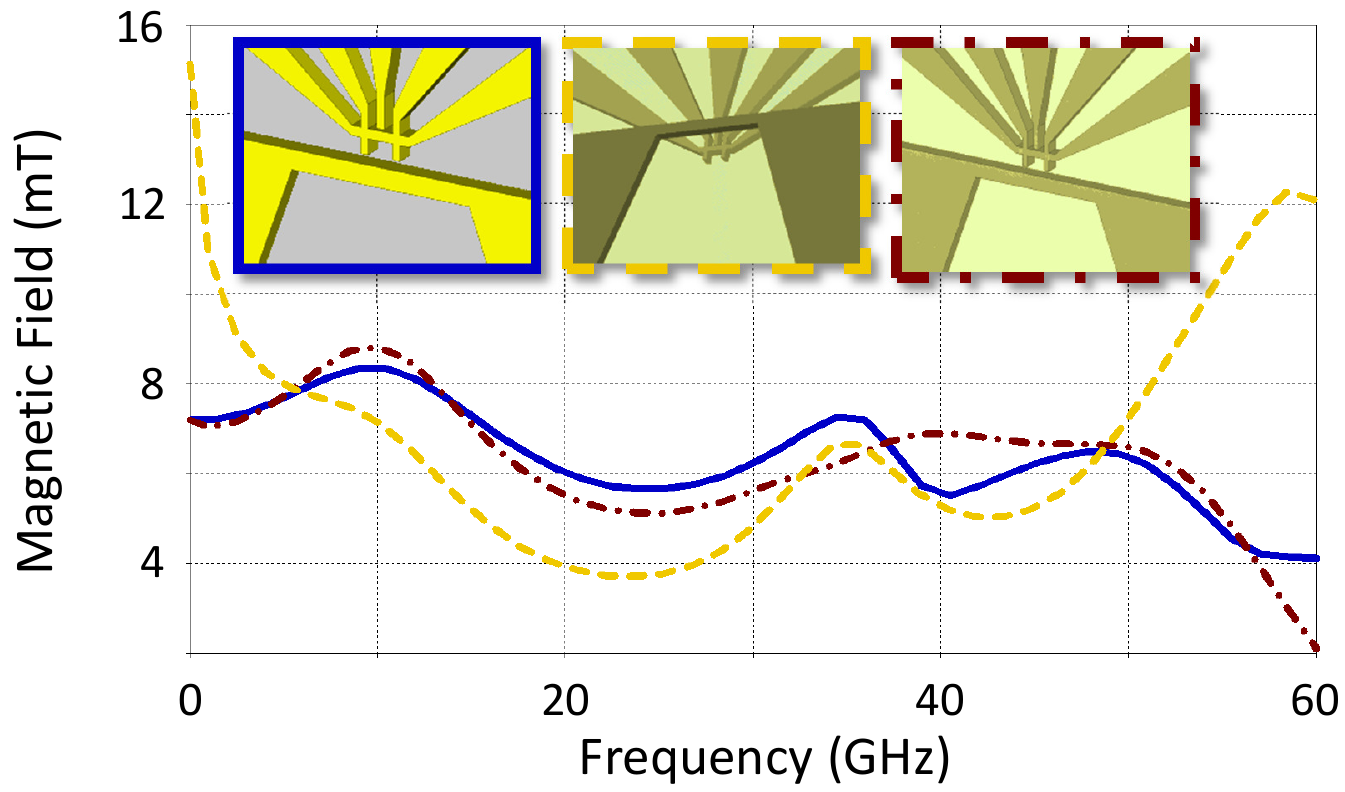}}\\
\subfloat{\includegraphics[width=0.5\textwidth]{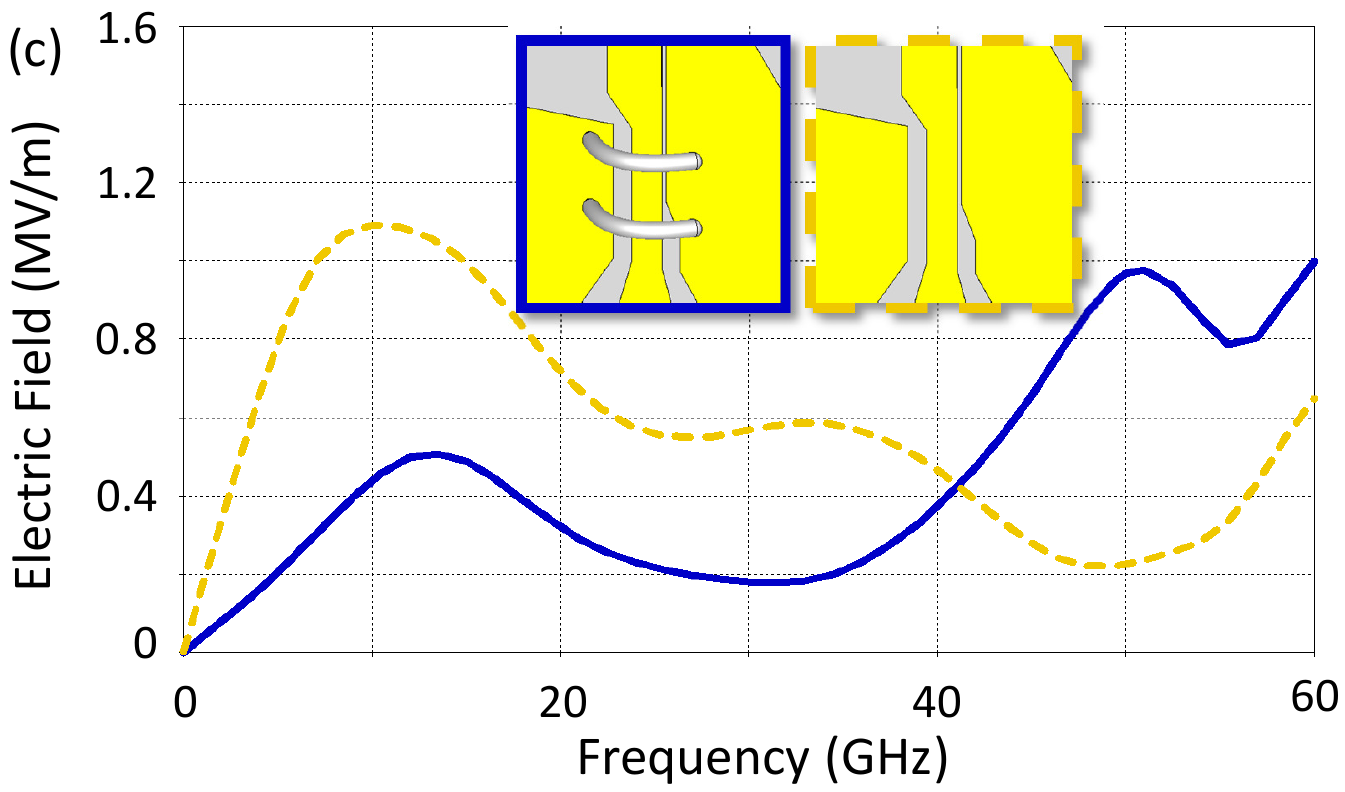}}
\subfloat{\includegraphics[width=0.5\textwidth]{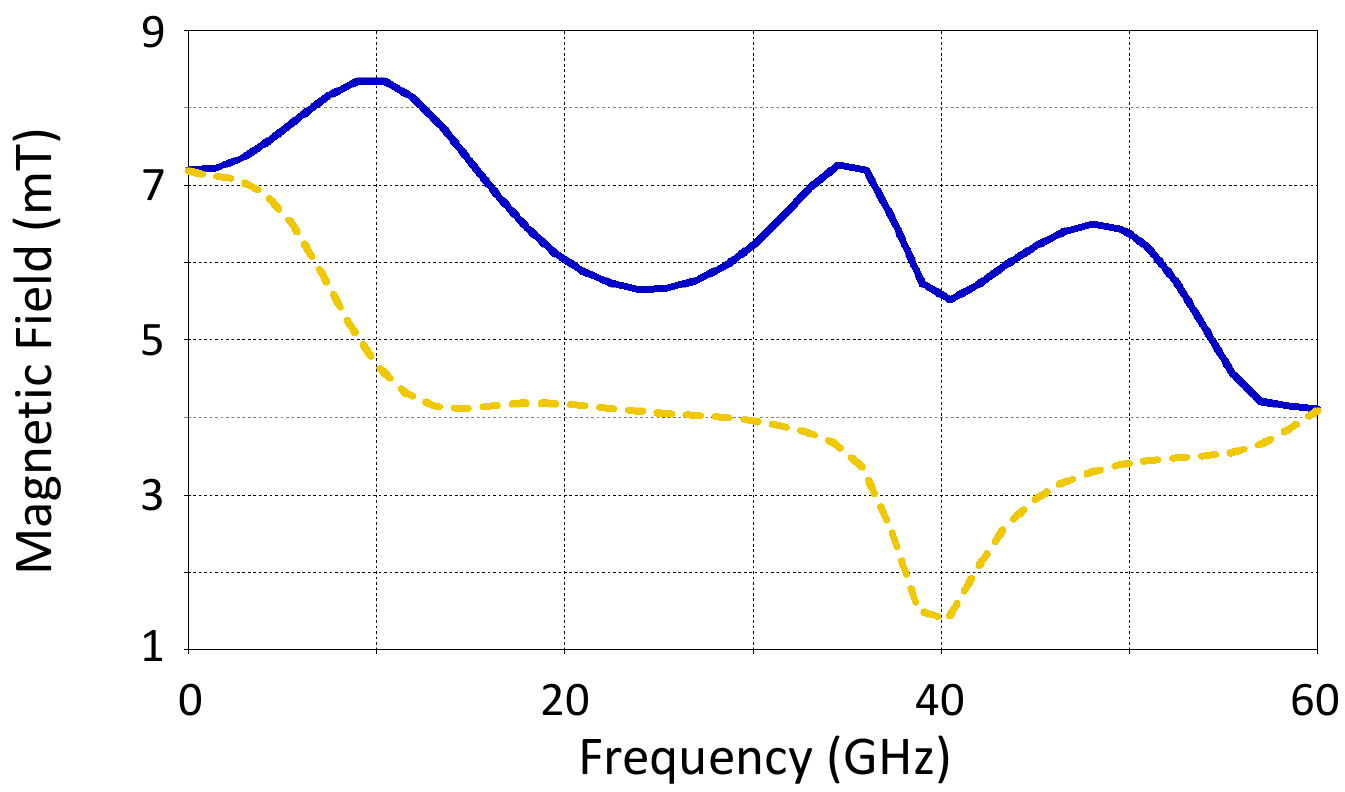}}
\end{center}
\caption{(a)~Comparison of S$_{11}$ parameters for different bond wire lengths and types:
200~$\mu$m bond wire (\full); 500~$\mu$m bond wire (\dashed);
150~$\mu$m bond wire (\dotted); 200~$\mu$m ribbon bond 70~$\mu$m
wide by 10~$\mu$m thick (\chain). (b)~Comparison between: the standard
model with the qubit located outside the loop (\full); model with
the transmission line elevated with calixarene, with the qubit
located inside the loop (\dashed); standard model with a thin film
of calixarene surrounding the transmission line and gates (\chain). (c)~Electric and magnetic fields obtained from the model with (\full) and without (\dashed) the bond wires bridging the ground plates of the on-chip CPW.
The coplanar structure under study on all subfigures is the on-chip balun shown in \fref{fig:ChipBalun}. The device and probe locations with respect to the substrate are
identical in all models.}\label{fig:Optimization}
\end{figure}

As explained in section~\ref{sec:Results}, the discontinuity between
the PCB and the chip causes a stub resonance at high frequency. Bond
wires have an effective inductance that increases with length and
decreases with thickness~\cite{gupta81}. In \fref{fig:Optimization}a we show that with very long bond wires the
resonance is very pronounced at the frequency calculated in
section~\ref{sec:Results}. Shorter bond wires push the resonance to higher frequencies. This behaviour suggests that the bond wires can be modelled as an
inductance in parallel with the R-L-C equivalent circuit of the
on-chip resonating stub. Therefore, minimizing the effective
inductance from the bond wires helps remove the resonant effect.

Using very short bond wires can completely
remove the resonant effect, but bond lengths of less than
200~$\mu$m can be challenging to realize. A more viable solution
is to use ribbon bonds, which have a much smaller effective
inductance than standard bond wires.\\

We have seen in  \fref{fig:2DEfield} that the electric field is mainly radiated towards the outside of the loop. Therefore, placing the spin readout device on the inside of the loop should decrease the electric field observed by the qubit, while maintaining high oscillating magnetic fields. In the experiment performed by Koppens \emph{et al.}~\cite{koppens06nat}, the SRD was indeed positioned on the inside of the loop, by separating the transmission line and the device with a thin film of dielectric. We can adjust our model to show results for this setup. We model a 100~nm coating of a material with $\epsilon_{eff} = 7.1$ (i.e. calixarene~\cite{holleitner03apl}) on top of the Si substrate and lay the planar loop on top, with the magnetic probe 50~nm inside the loop. We show in \fref{fig:Optimization}b the simulated $B_1$ and electric field amplitudes of the shorted on-chip balun with and without the dielectric. The results suggest that using the dielectric layer to place the device inside the loop does improve the magnetic to electric field ratio at frequencies below 20~GHz and above 40~GHz. We also find that by adding the dielectric alone, without modifying the position of the loop, a shielding effect causes a reduction of the electric field at the SRD, without affecting the magnetic field.\\

We explained at the end of \sref{sec:Designs} the importance of bridging the ground plates of a CPW at discontinuities, in order to suppress the excitation of even modes of propagation. The simulation results in \fref{fig:Optimization}c show how
removing the bond wires that bridge the ground planes leads to a some deterioration of the frequency response.

\section{Benchmark with experiments} %%%%%%%%%%%%%%%%%%%%%%%%%%%%%%%%%%%%%%%%%%%%%

\begin{figure}
\begin{center}
\subfloat{
\includegraphics[width=0.45\textwidth]{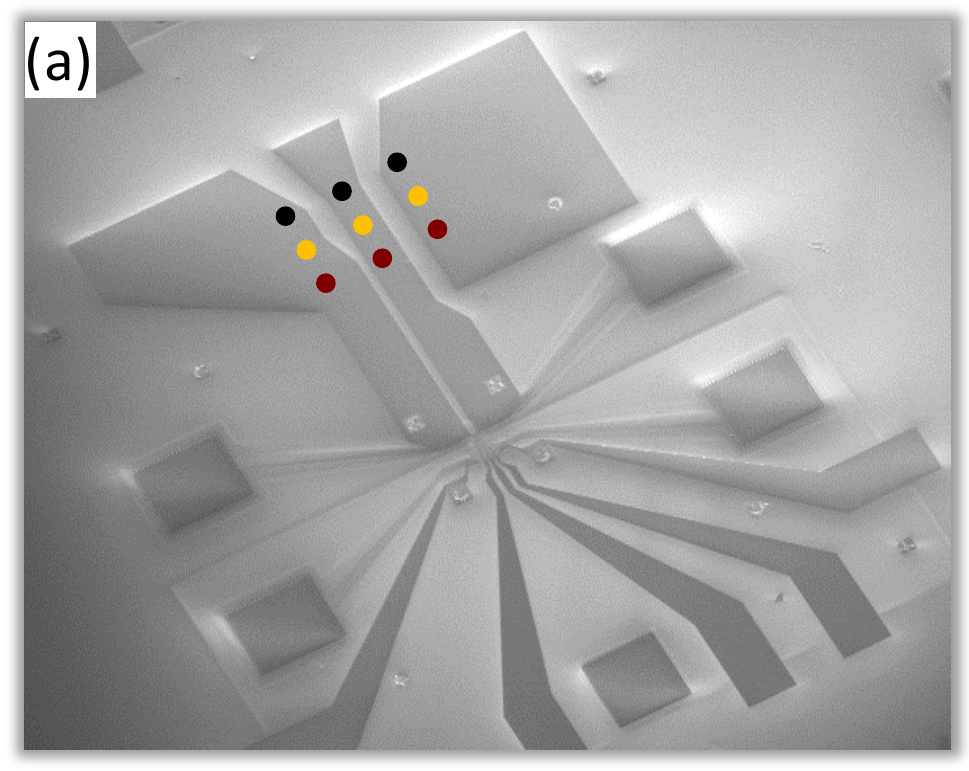}\label{fig:Probes}}
\subfloat{
\includegraphics[width=0.55\textwidth]{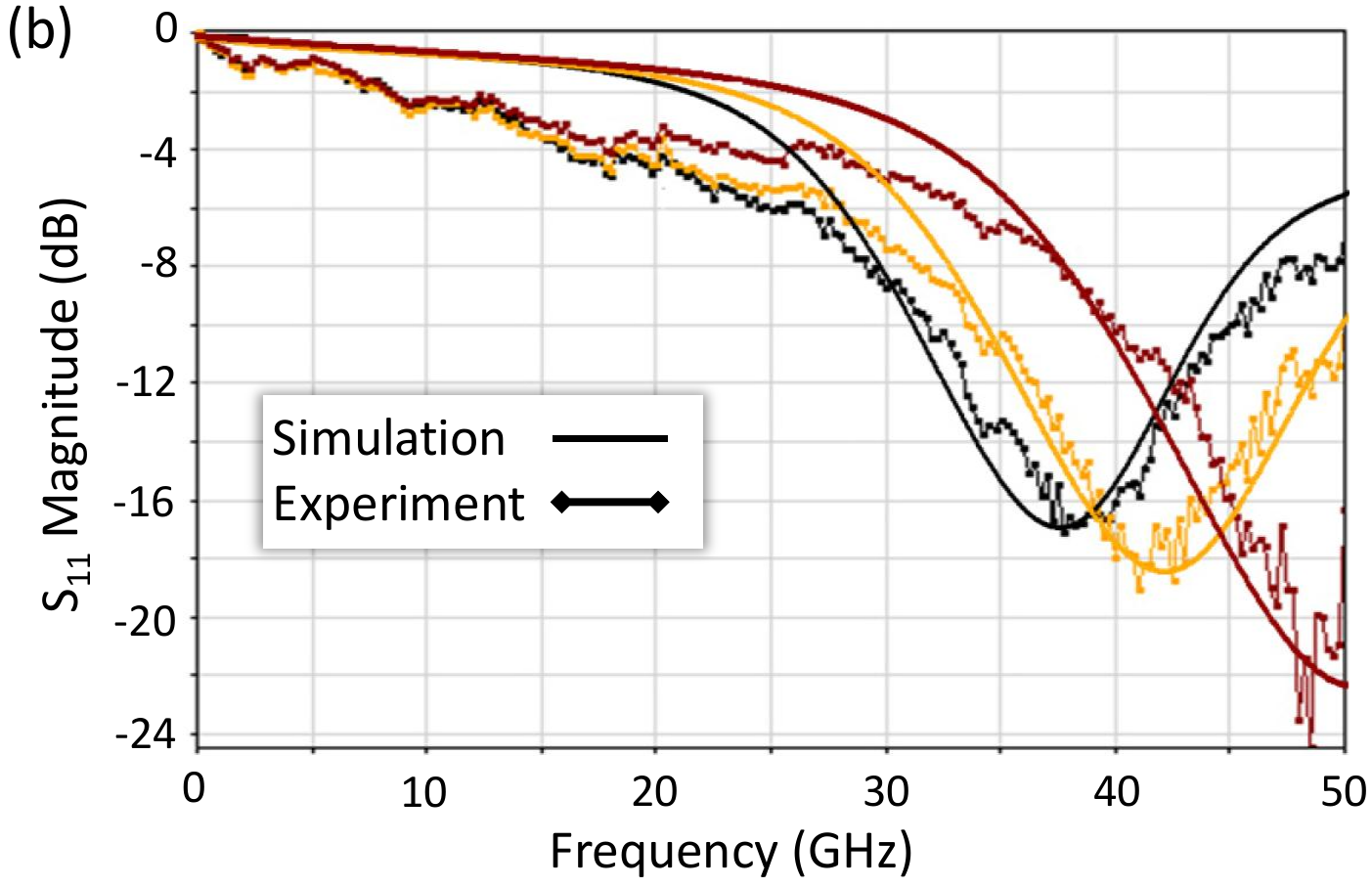}\label{fig:S11Exp}}
\caption{(a)~Scanning electron microscope image of a spin readout device coupled to a microwave transmission line with on-chip balun as sketched in \fref{fig:ChipBalun}. Coloured dots show microwave probe locations corresponding to each S$_{11}$ result. (b)~Simulated (solid lines) and experimental (dots) S$_{11}$ parameters for the on-chip balun.}
\end{center}
\end{figure}

\begin{figure}
\begin{center}
\subfloat{
\includegraphics[width=0.47\textwidth]{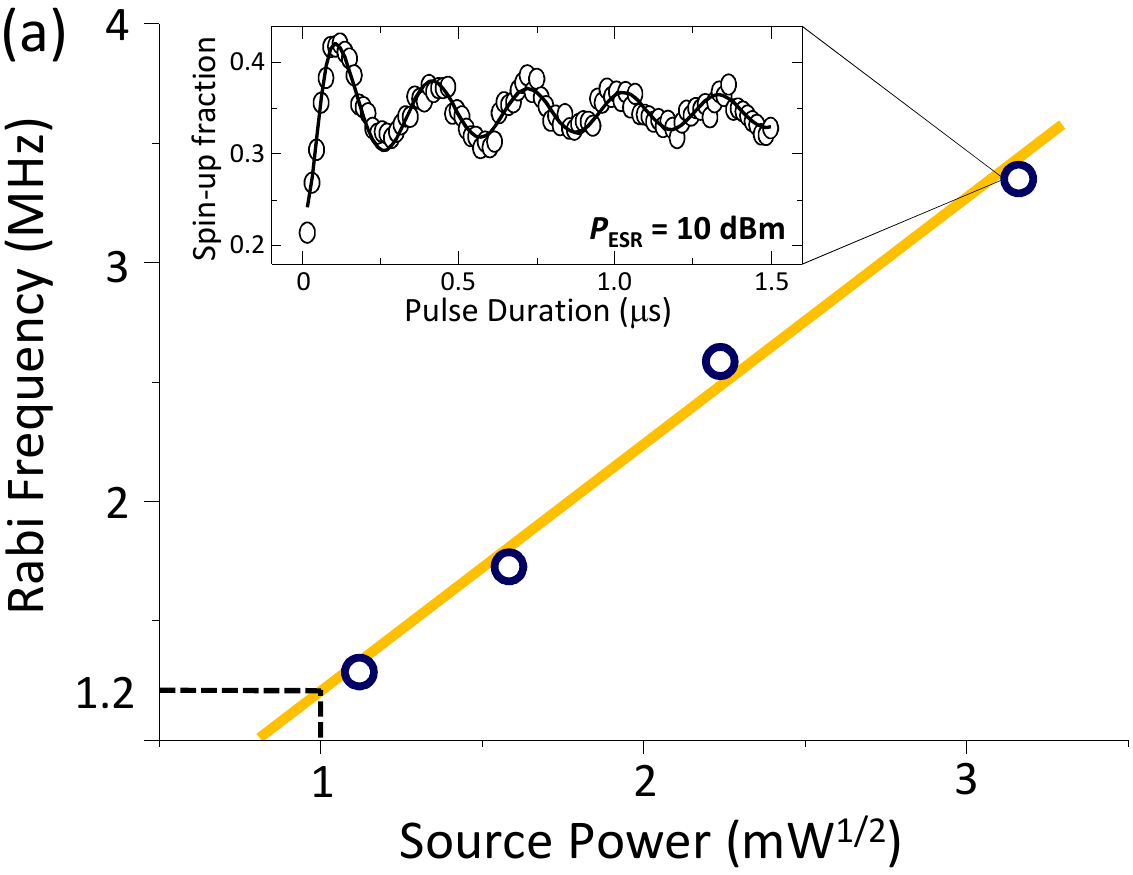}\label{fig:Rabi}}
\subfloat{
\includegraphics[width=0.53\textwidth]{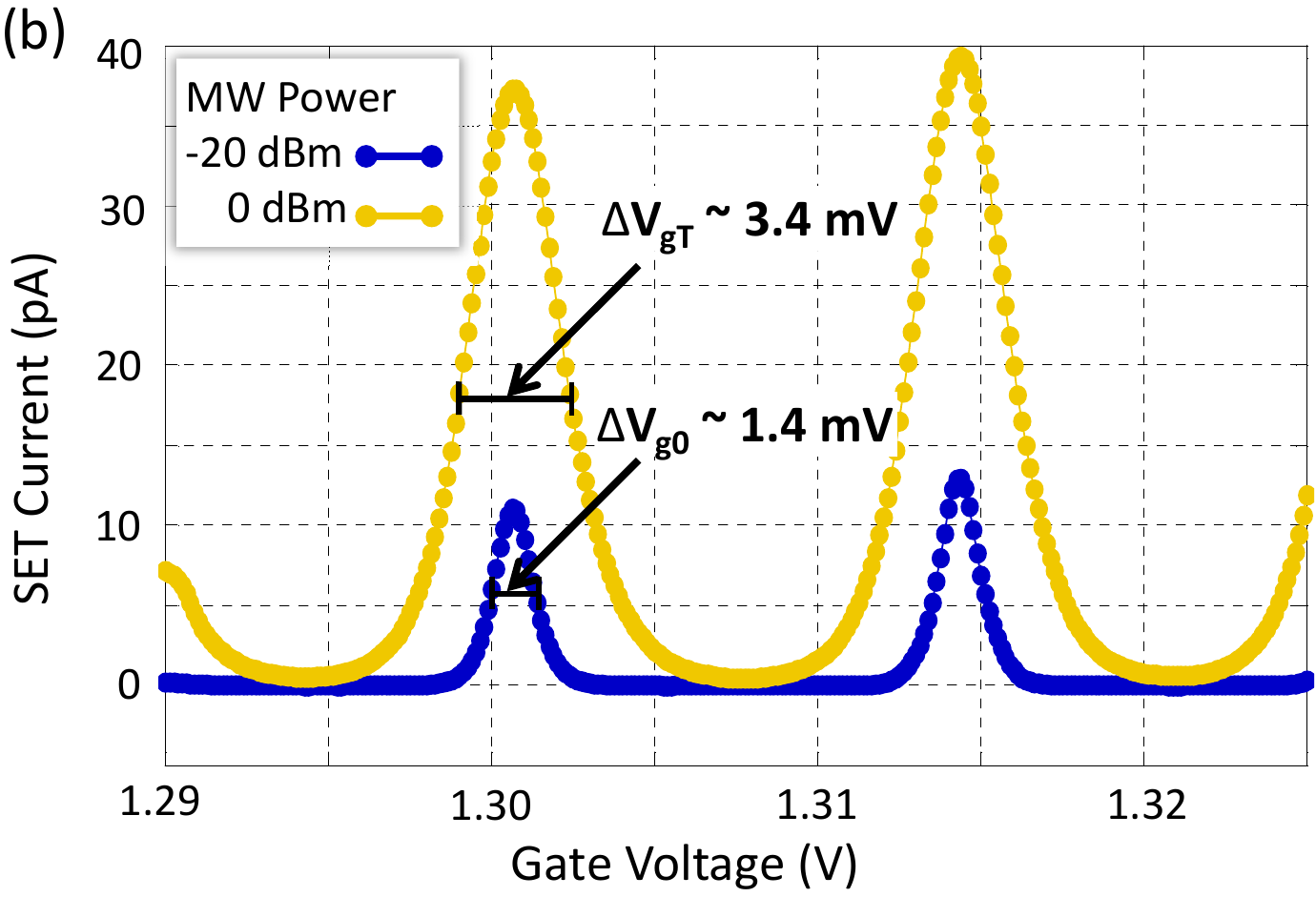}\label{fig:Peaks}}
\caption{(a)~Power-dependence of the electron Rabi frequency for pulsed ESR measurements performed at 30~GHz. $B_1$ is extracted from the linear fit of the Rabi frequency at 1 mW$^{1/2}$ highlighted in the figure. Inset shows a sample electron spin Rabi oscillation measured at 10 dBm of power from the microwave source. Data taken from Pla \emph{et al.}~\cite{pla12nat}. (b)~SET current as a function of gate voltage, in the presence of a continuous-wave microwave excitation at 30 GHz. The data was taken on the same device used by Pla \emph{et al.}~\cite{pla12nat}. Notice the broadening of the Coulomb peaks at high microwave power. The microwave powers quoted in the figure are the output powers at the source. The power at the chip is approximately 30 dB lower due to loss along the coaxial line. The electric field radiated by the loop is extracted from the width of each of the peaks labelled on the figure.} \end{center}
\end{figure}

We have fabricated and operated a complete spin qubit structure as sketched in \fref{fig:Design} and shown in \fref{fig:Probes}. Due to the difficulty in making wire bonds of such short lengths, we were not able to bridge the ground plates of the on-chip balun in the fabricated device.

\subsection{Stub resonance of the on-chip CPW/CPS balun} %%%%%%%%%%%%%%%%%%%%%%%%%%%%%%%%%%%%%%%%%%%%%

As a first experimental benchmark of the reliability of the models discussed above, we made an S$_{11}$ measurement using a 10~MHz to 50~GHz Agilent PNA microwave network analyzer. Instead of bonding the planar line to a PCB, we have used a microwave probe station, which terminates in a 3-terminal probe that can be placed directly on the chip at various locations. This allows us to study the stub resonance of the on-chip balun structure, by varying the distance between the edge of the chip and the contact point of the probe. To simulate this setup accurately, we removed the PCB structure from the model shown in \fref{fig:ChipBalun} and connected bond wires from the coaxial port to the on-chip line, at three different microwave probe locations. \Fref{fig:S11Exp} shows a very good agreement between the measured and simulated S$_{11}$ values.

\subsection{Oscillating magnetic field at the spin qubit}\label{sec:BField} %%%%%%%%%%%%%%%%%%%%%%%%%%%%%%%%%%%%%%%%%%%%%

The on-chip balun planar loop design shown in \fref{fig:Design} has been successfully employed to demonstrate coherent operation of a single-donor spin qubit in silicon~\cite{pla12nat}. In this work, Rabi oscillations were observed as a result of pulsed ESR experiments on the donor electron (inset \fref{fig:Rabi}). The frequency of the Rabi oscillations is given by $f_{Rabi} = (g\mu_BB_1)/h$. The fit from the power-dependence in \fref{fig:Rabi} shows $f_{Rabi} = 1.2$ MHz at 0~dBm power from the source, yielding an oscillating magnetic field at the donor $B_1 = 43$~$\mu$T. This experiment was performed at a spin resonance frequency $\nu_{ESR} = 30$ GHz.

In these experiments, the loss of the coaxial line connecting the source to the device was $\sim30$~dB at 30 GHz, measured independently. Rescaling the simulations results in \fref{fig:Optimization}c to a power of -30~dBm at the PCB, and accounting for the rotating wave approximation (\sref{sec:Guidelines}), our simulations predict 63~$\mu$T of $B_1$ at the donor (\fref{fig:B1Sim}). We consider this to be a very good agreement between the simulations and the experiment, and highlight the fact that the microwave simulation is quite successful in predicting surface currents and magnetic fields, even in a structure that shrinks to sub-micron dimensions.

\subsection{Electric field radiated to the spin readout device}\label{sec:EField} %%%%%%%%%%%%%%%%%%%%%%%%%%%%%%%%%%%%%%%%%%%%%

The use of an electrostatically-induced SET~\cite{angus07NL} as the spin readout device allows us to estimate the magnitude of the electric field produced by the planar loop at the location of the SET island~\cite{beenakker91prb}. The current through the SET as a function of gate voltage exhibits characteristic Coulomb peaks. The width of these peaks in units of gate voltage ${\Delta}V_g$ is a function of temperature, source-drain voltage bias and transparency of the potential barriers around the SET island. Because the SET island has a floating potential, an oscillating electric field $E_l$ produced by the loop will contribute an additional Coulomb peak broadening ${\Delta}V_{gE}$. A simple parallel-plate capacitor model between the gate and the SET island yields the relation $E_l = {\Delta}V_{gE}/d$, where $d$ is the distance between gate and SET island. In this case, $d \approx 8$~nm represents the thickness of the insulating SiO$_2$ layer.

A scan of the SET current as a function of the gate voltage shows broadening of the Coulomb peaks when a microwave excitation is applied to the loop (\fref{fig:Peaks}). The peak broadening ${\Delta}V_{gE}$ can be extracted from \fref{fig:Peaks} by writing ${\Delta}V_{gE} = \sqrt{{\Delta}V_{gT}^2 - {\Delta}V_{g0}^2}$, where ${\Delta}V_{gT}$ (${\Delta}V_{g0}$) is the width of the Coulomb peak with (without) the microwave excitation. This calculation yields $E_l \approx 380$~kV/m from the data in \fref{fig:Peaks}, taken at 30~GHz and 0~dBm power at the source.

Rescaling the results in \fref{fig:Optimization}c for a microwave source to device loss of $\sim30$~dB (see \sref{sec:BField}), we obtain a prediction of $E_l \approx 20$~kV/m. This large discrepancy suggests that there could be additional mechanisms causing Coulomb peak broadening when the microwave excitation is applied, and/or that the simulation does not capture all the factors that influence the electric field patterns. For instance, additional thermal broadening of the Coulomb peaks could arise from radiated heat due to the sizeable current flowing through the loop. Additional electric fields could arise from cavity modes excited in the metallic enclosure used in this experimental setup.

\section{Conclusions} %%%%%%%%%%%%%%%%%%%%%%%%%%%%%%%%%%%%%%%%%%%%%

In conclusion, we have presented an extensive study on the integration of microwave transmission lines in nanostructures, for the purpose of optimizing the design of magnetic resonance experiments aimed at single spins.

We have discussed a variety of topologies and explained the pros and cons of each one, supported by microwave simulation techniques that are specifically suited to deal with dimensions shrinking from millimetre to nanometre scale. We have then described a new structure that combines a good control of  microwave transmission modes with a maximized value of the magnetic field available to drive coherent control of a spin qubit. This structure has been employed to achieve coherent control of a single-donor electron spin qubit in silicon. The experiments show that the simulation describes accurately the magnetic field produced by the microwave loop, whereas the electric field modelling remains in need of better understanding.

We expect the design guidelines presented in this paper to be useful in assisting and facilitating the intense efforts towards performing coherent spin control in nanostructures, where high values of oscillating magnetic fields at frequency $\gg 10$~GHz must coexist with charge-sensitive devices.

\ack This research was conducted by the Australian Research Council
Centre of Excellence for Quantum Computation and Communication
Technology (project number CE11E0096), and supported in part by the US Army Research Office (W911NF-08-1-0527). We acknowledge access to the Australian National Fabrication Facility, and thank R. Ramer and K. Y. Chan for the use of the PNA and microwave probe station. We also thank J. T. Muhonen and A. Laucht for their feedback on the text and support in measurements.

\section*{References}
\bibliographystyle{unsrt}
\bibliography{QC_esr}

\end{document}